\documentclass[11pt]{article}
\usepackage{comment,soul}
\usepackage[hidelinks]{hyperref}
\usepackage{enumitem}
\usepackage{amsmath,amssymb,epsf,cite,graphicx,subfigure}
\usepackage{amsmath,braket,tikzsymbols}
\usepackage[bbgreekl]{mathbbol}
\usepackage{comment}
\usepackage[export]{adjustbox}
\usepackage{dsfont}
\usepackage{faktor}

\numberwithin{equation}{section}

\definecolor{airforceblue}{rgb}{0.36, 0.54, 0.66}
\newcommand{\beq}{\begin{equation}}
\newcommand{\eeq}{\end{equation}}

\begin{document}
\baselineskip=15.5pt
\pagestyle{plain}
\setcounter{page}{1}

\begin{center}
{\LARGE \bf Isometric evolution in de Sitter quantum gravity}
\vskip 1cm

\textbf{Jordan Cotler$^{1,a}$ and Kristan Jensen$^{2,b}$}

\vspace{0.5cm}

{\it ${}^1$ Society of Fellows, Harvard University, Cambridge, MA 02138, USA \\}
{\it ${}^2$ Department of Physics and Astronomy, University of Victoria, Victoria, BC V8W 3P6, Canada\\}

\vspace{0.3cm}

{\tt  ${}^a$jcotler@fas.harvard.edu, ${}^b$kristanj@uvic.ca\\}

\medskip

\end{center}

\vskip1cm

\begin{center}
{\bf Abstract}
\end{center}
\hspace{.3cm} 
We study time evolution in two simple models of de Sitter quantum gravity, Jackiw-Teitelboim gravity and a minisuperspace approximation to Einstein gravity with a positive cosmological constant. In the former we find that time evolution is isometric rather than unitary, and find suggestions that this is true in Einstein gravity as well.  The states that are projected out under time evolution are initial conditions that crunch. Along the way we establish a matrix model dual for Jackiw-Teitelboim gravity where the dilaton varies on the boundary.

\newpage

\tableofcontents

\section{Introduction}

Do the postulates of quantum mechanics survive in quantum gravity? The main tools for studying quantum gravity, the gravitational path integral (including its Hamiltonian formulation), and string theory, naturally produce states and transition amplitudes and so start by assuming most of the postulates. However the probabilistic interpretation of amplitudes, enforced by the unitarity of time evolution, is not guaranteed within the path integral formulation and has to be checked.

We use the gravitational path integral and find a simple mechanism whereby a sum over smooth geometries leads to isometric rather than unitary evolution, which we demonstrate in a simple model of de Sitter quantum gravity. The basic result is that some states evolve into singular spacetime geometries with a crunch, and others to a bounce, and the former are projected out under evolution. Evolution acts unitarily on the ``code subspace'' of states that do not develop a crunch, while ``crunch'' states are projected out under evolution. In this way the Hilbert space of bulk states is smaller than the space of asymptotic states appearing in the de Sitter $S$-matrix. We find this to be true in Jackiw-Teitelboim (JT) gravity, an unrealistic but non-perturbatively soluble model of two-dimensional dilaton gravity~\cite{Jensen:2016pah, Maldacena:2016upp, Engelsoy:2016xyb, Saad:2019lba} that has been the subject of much recent work~\cite{Mertens:2022irh}. We also find evidence from a minisuperspace approximation that this is true in Einstein gravity. We then speculate how our results may generalize to more realistic models of quantum gravity.

Our findings are consistent with a recent proposal~\cite{Cotler:2022weg} that time evolution is isometric for quantum gravity in expanding cosmologies. In~\cite{Cotler:2022weg}, one of us gave general arguments for isometric evolution and provided examples with matter effective field theory in rigid curved spacetimes.  In this paper we give a proof of principle for the proposal in a simple model of dynamical gravity.

Our analysis of de Sitter JT gravity builds upon previous work (including our own)~\cite{Maldacena:2019cbz, Cotler:2019nbi, Cotler:2019dcj}. In~\cite{Cotler:2019dcj} we studied the $S$-matrix of JT gravity to leading order in a topological expansion and to all orders in the gravitational coupling. We considered asymptotic states corresponding to large closed universes with a fixed renormalized length and on which the dilaton of JT gravity is constant. In between such states we found the infinite time evolution operator $\widehat{\mathcal{U}}$ to be a projector. In this paper we make sense of this projector, and see that it is a consequence of isometric evolution.  We analyze the Hilbert space of de Sitter JT gravity at intermediate times, and find a basis in which we can cleanly identify states that correspond to bouncing and crunching cosmologies, as well as the change-of-basis matrix to the basis of asymptotic states. The sum over smooth geometries projects out the crunching states. We are able to write the infinite evolution operator as $\widehat{\mathcal{U}} = \widehat{V}\widehat{V}^{\dagger}$ where $\widehat{V}$ is the evolution operator from a bulk time to the infinite future (with $\widehat{V}^{\dagger}$ the evolution from the infinite past to a bulk time). Crucially, we find that $\widehat{V}$ is an isometry. Furthermore, while previous work on de Sitter JT gravity involved asymptotic states with constant dilaton, our analysis allows for arbitrary asymptotic states and we find isometric evolution in this richer setting. In particular, we find an infinity of null asymptotic states, so that asymptotic states with a varying dilaton differ from those with a constant dilaton by a null state. Along the way we find the dictionary between JT gravity with these boundary conditions and a double scaled matrix model.

We then go on to study time evolution in a minisuperspace approximation of Einstein gravity with a positive cosmological constant where the spatial universe is a round sphere. We treat this minisuperspace approximation quantum mechanically, with the result that the Hilbert space of bulk states corresponds to cosmologies which bounce or crunch. The latter are projected out by evolution, consistent with isometric rather than unitary evolution. We wrap up with a discussion, suggesting that a similar mechanism should apply to more realistic models of quantum gravity and perhaps to our own universe.


\section{de Sitter JT gravity}

JT gravity is a model of two-dimensional gravity with a dilaton $\phi$ and a metric $g$. The action of the de Sitter version is
\beq
	S_{\rm JT} =\frac{S_0}{4\pi} \int d^2x \sqrt{-g} \,R+  \int d^2x \sqrt{-g} \,\phi(R-2) +S_{\rm bdy}\,.
\eeq
The term proportional to $S_0$ is topological, and we take $S_0\gg 1$ to suppress fluctuations of the spacetime topology. The basic solution to the field equations is global dS$_2$ space,
\begin{align}
\begin{split}
\label{E:globaldS}
	ds^2 & = -dt^2 + \alpha^2\cosh^2(t)dx^2\,,
	\\
	\phi &= \phi_0 \sinh(t)\,,
\end{split}
\end{align}
where $x\sim x+2\pi$ and $\alpha$ is a modulus labeling the space. These are bounce cosmologies where the spatial universe is a circle reaching a minimum size of $2\pi |\alpha|$ in between two asymptotically dS$_2$ regions reached as $t \to \pm \infty$. The general definition of an asymptotically future dS$_2$ region is a line element and dilaton which behave as
\begin{align}
\begin{split}
\label{E:asyFuture}
	ds^2 & = -dt^2 + (e^{2t} + O(1)) dx^2\,,
	\\
	\phi & = \frac{1}{2\pi}\,e^{t+ \varphi(x)} + O(1)\,,
\end{split}
\end{align}
as $t\to\infty$, and similarly for a past asymptotic region. From this form we define asymptotically future dS$_2$ boundary conditions as follows. We introduce a boundary at $t=\ln \Lambda$ with $\Lambda$ tending to infinity, on which the induced metric is $\sim\Lambda^2dx^2$ and the dilaton is $\sim\Lambda e^{\varphi(x)}/2\pi$. We add the boundary term
\beq
\label{E:Sbdy}
	S_{\rm bdy} =-\frac{S_0}{2\pi}\int_{\partial M}dx\sqrt{\gamma}\,K -2 \int_{\partial M} dx \sqrt{\gamma} \,\phi(K-1)\,,
\eeq
to the action, with $\gamma$ the induced metric and $K$ the extrinsic curvature of the boundary, and take the boundary to infinity. Through this procedure we fix a large future boundary with a renormalized boundary metric $dx^2$ and a renormalized dilaton $e^{\varphi(x)}$. The boundary term in the action is required so that JT gravity has a consistent variational principle with these boundary conditions.

This boundary is spacelike and therefore prepares a final quantum state labeled by the dilaton profile. We notate a final state with dilaton profile $e^{\varphi_1}$ as $\langle e^{\varphi_1}|$. Similarly, asymptotically past dS$_2$ boundary conditions with a dilaton profile $e^{\varphi_2}$ prepare an initial quantum state that we notate as $|-e^{\varphi_2}\rangle$.\footnote{The relative minus sign in our labeling of past and future asymptotic states can, for now, be understood as a convention, having to do with the fact that asymptotic states are actually characterized by the renormalized dilaton times the sign of the extrinsic curvature of the boundary circle. We further comment on it in Appendix~\ref{App:details2}.}  Because these states are prepared in the far past and future we call them \emph{asymptotic} states.\footnote{There are also multi-universe asymptotic states where the initial or final space is a disjoint set of $n$ large circles, each of which is characterized by a renormalized dilaton.} 

Previous work~\cite{Maldacena:2019cbz, Cotler:2019nbi, Cotler:2019dcj, Moitra:2022glw} on de Sitter JT gravity focused on asymptotic states where the dilaton is constant, and so is incomplete since the most general asymptotic state has a varying dilaton. Even so, the three main quantities considered were: (1) the wavefunction at future infinity of the no-boundary (Hartle-Hawking) state $|\varnothing\rangle$ of de Sitter JT gravity, where there is no past and the future is a large asymptotic circle; (2) the sum over spacetimes with the topology of global dS$_2$, comprising the infinite-time transition amplitude between an asymptotic circle in the past and an asymptotic circle in the future; and (3) the inner product on asymptotic states~\cite{Cotler:2019dcj}. The inner product is required to obtain properly normalized transition amplitudes. (In our previous work~\cite{Cotler:2019dcj} we also proposed a topological expansion for de Sitter JT gravity which we do not discuss in the present work.) 

We proceed to study the more general asymptotic states. We relegate a detailed description to the Appendix. The main point is that JT gravity has no bulk degrees of freedom; the dilaton acts as a Lagrange multiplier enforcing the constant curvature condition, uniquely fixing the spacetime metric up to moduli that must be integrated over with the correct measure. Furthermore, as in the AdS version of JT gravity, each asymptotic boundary is equipped a single boundary degree of freedom, a ``Schwarzian mode,'' which has to be integrated over in the quantum theory. The action for the Schwarzian mode is more complicated when the dilaton varies, but as we explain in the Appendix its path integral can still be computed exactly. We note that the action is only sensible when $e^{\varphi}$ is everywhere positive, or everywhere negative, and formally we must equip $e^{\varphi}$ with an infinitesimal imaginary part.

\begin{figure}
\centering
\includegraphics[scale=.6]{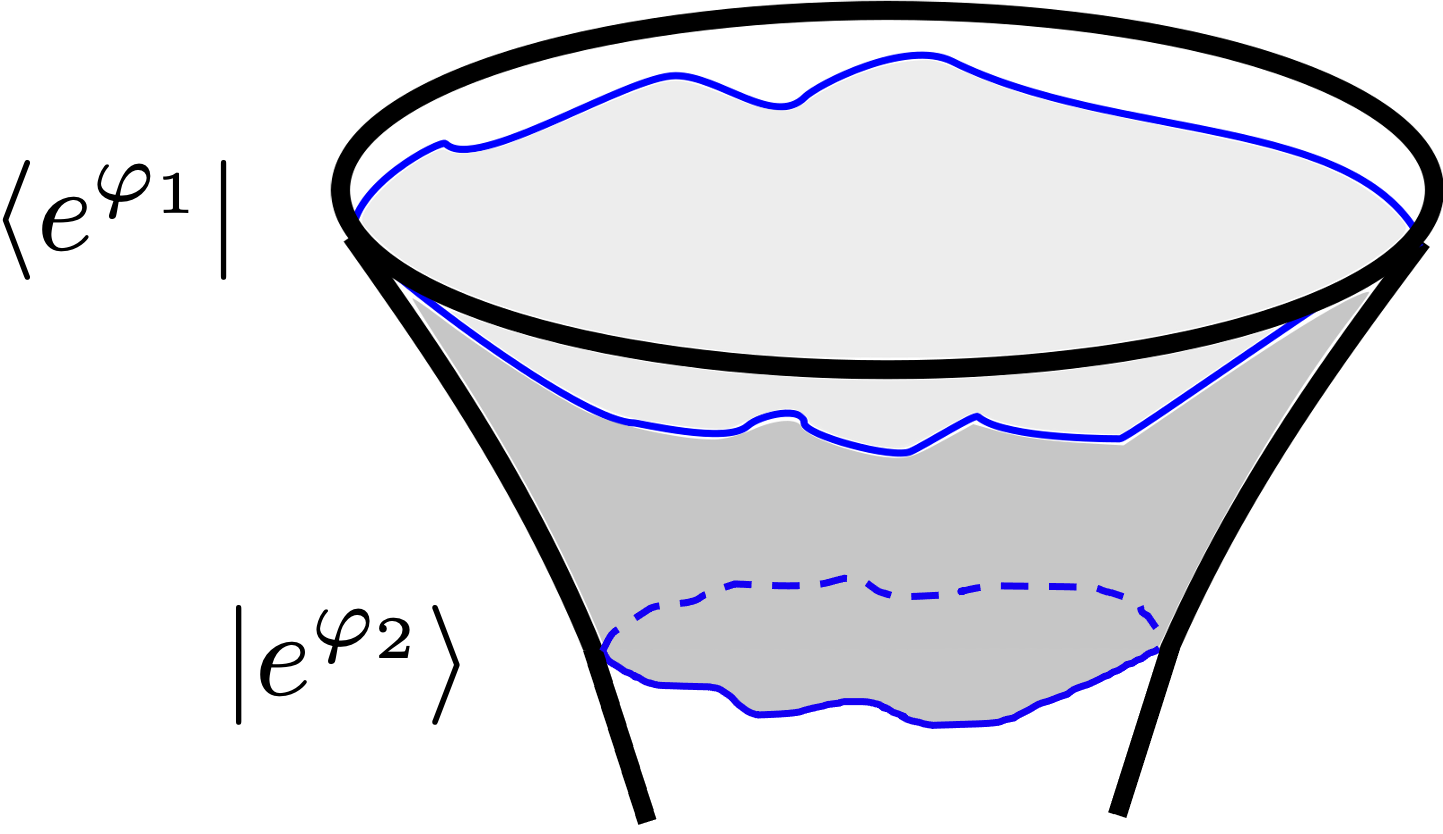}
\caption{A depiction of the inner product $\langle e^{\varphi_1}|e^{\varphi_2}\rangle$.  Following~\cite{Cotler:2019dcj}, we consider boundary conditions in the future asymptotic region corresponding to a bra and a ket, and perform the path integral over those metrics that interpolate between the boundary conditions in the limit that the corresponding boundaries approach one another.\label{Fig:IP1}}
\end{figure}

Following the methods of~\cite{Cotler:2019dcj} and accounting for the Schwarzian path integral with a varying dilaton, we find that the inner product of the asymptotic states $|e^{\varphi_1}\rangle$ and $|e^{\varphi_2}\rangle$ is
\beq
\label{E:asymptoticIP1}
	\langle e^{\varphi_1}|e^{\varphi_2}\rangle = \sqrt{\Phi_1 \Phi_2} \,\delta(\Phi_1-\Phi_2) \,e^{i\mathcal{S}[\varphi_1]-i \mathcal{S}[\varphi_2]}\,,
\eeq
where we define $\mathcal{S}[\varphi] := \frac{1}{2\pi}\int_0^{2\pi}dx\,e^{\varphi} \varphi'(x)^2$ and also $\Phi_i^{-1} := \frac{1}{2\pi}\int_0^{2\pi} dx\,e^{-\varphi_i}$. A depiction of the inner product, along the lines of~\cite{Cotler:2019dcj}, can be seen in Figure~\ref{Fig:IP1}. The result~\eqref{E:asymptoticIP1} implies an infinite redundancy in the spectrum of asymptotic states. Consider two dilaton profiles $e^{\varphi_1}$ and $e^{\varphi_2}$ with the property that $ \Phi_1=\Phi_2$. Then all states of the form
\beq
	\ket{\Psi} = e^{i \mathcal{S}[\varphi_1]}|e^{\varphi_1}\rangle - e^{i \mathcal{S}[\varphi_2]} |e^{\varphi_2}\rangle
\eeq
are null. We construct a physical Hilbert space $\mathcal{H}_{\rm asy}$ in the usual way by identifying any two states that differ by a null state. Under that identification the state $|e^{\varphi}\rangle$ is identified with $e^{-i S[\varphi]} |\Phi\rangle$, i.e.~the state characterized by a constant dilaton with the same Fourier zero mode as $e^{-\varphi}$. The physical Hilbert space is then spanned by equivalence classes whose representatives are states with a constant dilaton $|\Phi\rangle$, with inner product $\langle \Phi_1|\Phi_2\rangle = \sqrt{\Phi_1\Phi_2} \,\delta(\Phi_1-\Phi_2)$. In de Sitter JT gravity the dilaton can be positive or negative and $\Phi \in \mathbb{R}$. So the Hilbert space of asymptotic states $\mathcal{H}_{\rm asy}$ is isomorphic to the Hilbert space of a quantum mechanical particle on the line. We then rescale asymptotic states as $|\Phi\rangle \to \frac{|\Phi\rangle}{\sqrt{\Phi}}$ so that they have the standard inner product $\langle \Phi_1|\Phi_2\rangle = \delta(\Phi_1-\Phi_2)$, and so have a completeness relation $\mathds{1} = \int d\Phi |\Phi\rangle\langle \Phi|$.

Note that this analysis implies that the work of~\cite{Cotler:2019dcj} with constant dilaton states was complete after all. 

The existence of the null states is a consequence of large diffeomorphisms. Consider an asymptotically future dS$_2$ region~\eqref{E:asyFuture}. There is a family of diffeomorphisms that preserve the form of the line element and dilaton, but change the renormalized dilaton. These diffeomorphisms are ``large,'' acting all the way to the boundary, and so we do not divide by them in the sum over metrics. However, they relate asymptotic states. To be more precise, the transformation
\begin{align}
\begin{split}
	e^t & \to \frac{e^t}{x'(y)}\left( 1 - \frac{e^{-2t}}{4}\frac{x''(y)^2}{x'(y)^2} + O(e^{-4t})\right)\,,
	\\
	x & \to x(y) +\frac{e^{-2t}}{2}\,x''(y) + O(e^{-4t})\,,
\end{split}
\end{align}
preserves~\eqref{E:asyFuture} while acting on the renormalized dilaton as
\beq
	e^{-\varphi(x)} \to x'(y) \, e^{-\varphi(x(y))}\,.
\eeq
Here $x(y)$ is a reparameterization of the spatial circle obeying $x(y+2\pi)=x(y)+2\pi$ and $x'(y)\geq 0$. This transformation preserves the Fourier zero mode of $e^{-\varphi}$, namely $\Phi^{-1}$.  Moreover, because the dilaton has to be everywhere nonzero, the transformation can be used to relate any two dilaton profiles with the same zero mode $\Phi$. In particular, once we know from our Schwarzian analysis that the dilaton must be either always positive or always negative, then we learn that there is a large diffeomorphism that relates the state $\langle e^{\varphi}|$ to the constant dilaton state $\langle \Phi|$.

\begin{figure}[t]
\centering
\includegraphics[scale=.42]{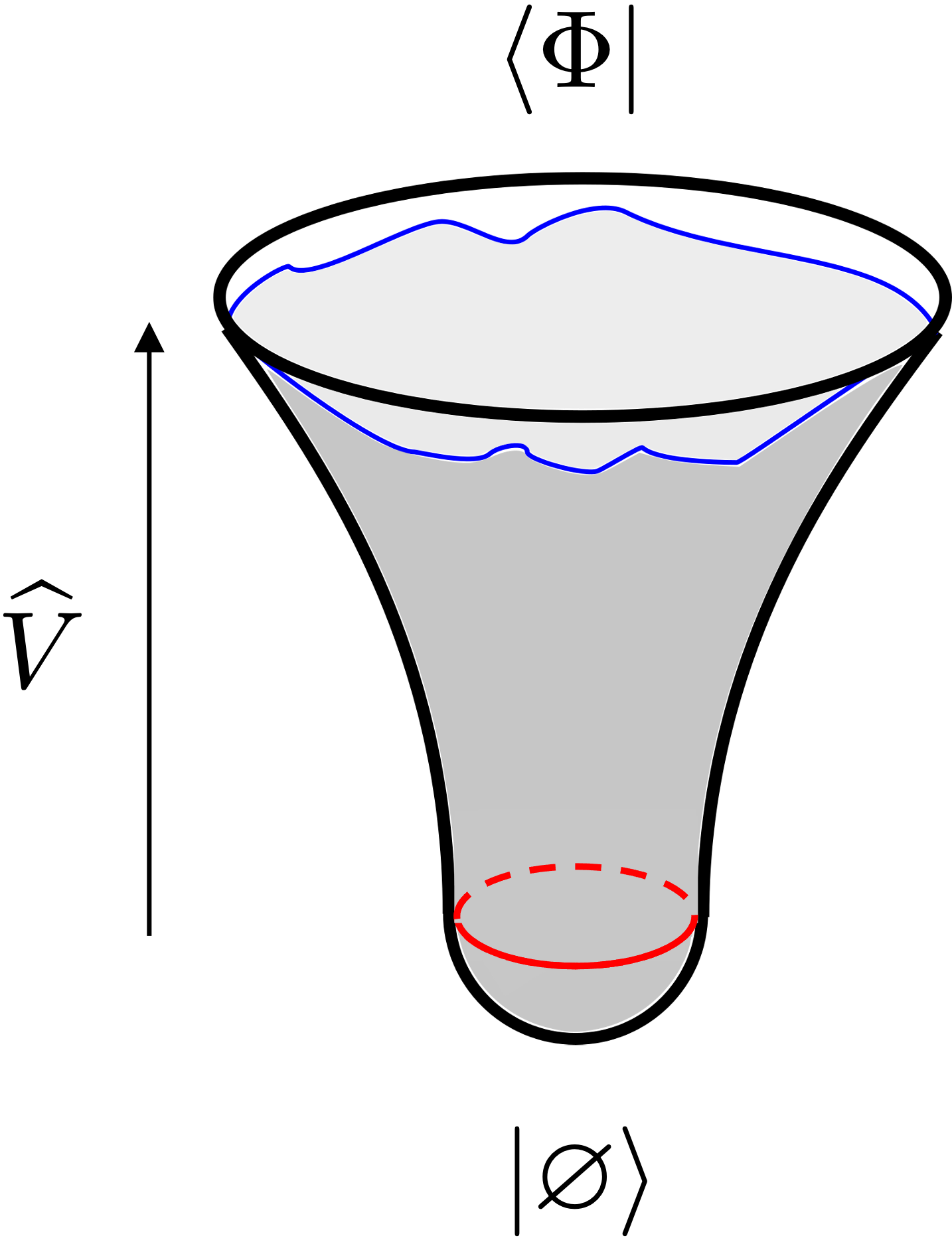}
\caption{The no-boundary state evolved to the infinite future to give the Hartle-Hawking state.  The state $|\varnothing\rangle$ corresponding to the Euclidean cap is prepared at a finite time, and is then evolved in Lorentzian time by $\widehat{V}$ to the infinite future.  The wavefunction is naturally computed in the $\Phi$-basis by projecting onto $\langle \Phi|$ in the far future.  \label{Fig:HH1}}
\end{figure}

In the basis of properly normalized asymptotic states the wavefunction of the no-boundary state at future infinity is, to leading order in the topological expansion,
\beq
	\Psi_{\rm HH}(\Phi) = \langle \Phi|\text{HH}\rangle = \langle \Phi|\widehat{V}|\varnothing\rangle  \approx  \frac{i^{\frac{3}{2}}\Phi}{\sqrt{2\pi}}  \,e^{ S_0+i \Phi}\,,
\eeq
which notably is non-normalizable: $\int d\Phi |\Psi_{\rm HH}(\Phi)|^2 \propto \int dx \,x^2$ diverges. Here $\widehat{V}$ is the semi-infinite evolution operator from the bulk time at which the no-boundary state is created to the infinite future.   A depiction is shown in Figure~\ref{Fig:HH1}.

Now consider the infinite-time transition amplitudes between asymptotic states with a large universe with $\Phi_2$ in the past and a large universe with $\Phi_1$ in the future. The result from~\cite{Cotler:2019dcj} for that amplitude, coming from the sum over cylinder geometries that smoothly connect the past and future circles, is
\beq
	\langle \Phi_1|\,\widehat{\mathcal{U}}\,|\Phi_2\rangle \approx \frac{i}{2\pi}\frac{1}{\Phi_1-\Phi_2+i \epsilon}\,,
\eeq
where $\approx$ means that we are neglecting higher order terms in the genus expansion, and where we have included an $i\epsilon$ prescription that renders the JT path integral convergent. The pole in this amplitude corresponds to the global dS$_2$ saddle~\eqref{E:globaldS} which, with our convention that past asymptotic states are labeled by $|-e^{\varphi}\rangle$, has $\Phi_1=\Phi_2$. Now consider a change of basis from states of definite $\Phi$ to those of its canonical conjugate, which we will call $p$ (thinking of $\Phi$ as a position and $p$ as a momentum). In the $p$-basis we have
\beq
	\langle p_1|\,\widehat{\mathcal{U}}\, |p_2\rangle\approx \Theta(p_1)\delta(p_1-p_2)\,.
\eeq
In this basis time evolution is very simple, and we see that infinite time evolution is unitary on the ``code subspace'' of states $|p\rangle$ with $p>0$.  In fact, there is a matrix model interpretation of this result which we found in our previous work~\cite{Cotler:2019dcj}; see Appendix~\ref{App:details2} for a short discussion of this.

In the next Section we will see that $p$-eigenstates are readily interpreted as bulk states, where $p>0$ correspond to bouncing cosmologies and $p<0$ to a crunch cosmologies.


\section{Isometric evolution in JT}

Now we consider bulk states. We find it convenient to fix a ``temporal gauge'' in which the line element reads
\beq
	ds^2 = -dt^2 +A(t,x)^2dx^2\,.
\eeq
With this gauge-fixing and on a finite-time cylinder\footnote{Any metric on a cylinder can be put into this form up to a large diffeomorphism, which can be understood to act on the initial and final states.} the JT action reads
\beq
	S =2 \int dt \,dx \,\phi\,(\ddot{A}-A)+ (\text{bdy})\,.
\eeq
Under the field redefinition $\mathcal{Q} = \frac{\phi}{\dot{A}}$ and $\mathcal{P} =A^2-\dot{A}^2$ the action is simply
\beq
	S = - \int dt \,dx\,\mathcal{Q} \dot{\mathcal{P}} + (\text{bdy})\,,
\eeq
an extremely simple quantum mechanics in Hamiltonian form. After this field redefinition we can adjust boundary terms so as to fix $\mathcal{Q}(x)$ on a constant time slice, which prepares a state $|\mathcal{Q}(x)\rangle$.  Alternatively, we can adjust boundary terms so as to fix $\mathcal{P}(x)$ on a constant time slice, which prepares a state $|\mathcal{P}(x)\rangle$. In an asymptotically de Sitter region $\lim_{t\to\infty}\mathcal{Q}(x) = e^{\varphi(x)}/2\pi$ and so asymptotic states with fixed renormalized dilaton correspond to $\mathcal{Q}$-eigenstates. On the other hand, we can fix the initial and final states to be $\mathcal{P}$-eigenstates. Asymptotically these are Neumann-like boundary conditions, but at finite time they naturally produce bulk states.

Integrating out $\mathcal{Q}$ enforces that $\mathcal{P}$ is conserved at each $x$. Because $\mathcal{P}(x)$ is conserved quantum mechanically we can deduce the corresponding metric, with
\beq
	A = c_+(x) \,e^t + c_-(x)\,e^{-t}\,.
\eeq
Requiring that the metric is everywhere smooth, which implies $c_+$ and $c_-$ are nonzero, there is a residual large diffeomorphism that ``straightens out'' $\mathcal{P}(x)$ so that it is a constant which we call $\mathcal{P}$. Initial states are equivalence classes labeled only by this constant which obey $\langle \mathcal{P}_1|\mathcal{P}_2\rangle = \delta(\mathcal{P}_1-\mathcal{P}_2)$. This puts $A^2$ into the form
\beq
	A^2 = \begin{cases} \mathcal{P} \cosh^2(t)\,,  & \mathcal{P}>0 \,, \\ |\mathcal{P}| \sinh^2(t) \,, & \mathcal{P}<0\,. \end{cases}
\eeq
The former is simply global dS$_2$ with $\mathcal{P}=\alpha^2$, while the latter is singular at $t=0$. So $\mathcal{P}>0$ states correspond to bounce cosmologies and $\mathcal{P}<0$ states to crunch cosmologies. The latter are projected out in the path integral formulation thanks to the sum over smooth geometries. That is, we build a bulk Hilbert space $\mathcal{H}_{\rm bulk}$ out of superpositions of $|\mathcal{P}\rangle$'s with $\mathcal{P}>0$. On that space finite time evolution $\widehat{U}$ simply acts as the identity, with
\beq
	\langle \mathcal{P}_1 |\,\widehat{U}\,|\mathcal{P}_2\rangle \approx \delta(\mathcal{P}_1-\mathcal{P}_2)\,.
\eeq

Taking stock, we have two Hilbert spaces in de Sitter JT gravity: (1) a space of asymptotic states $\mathcal{H}_{\rm asy}$ with a basis $|p\rangle$ with $p\in \mathbb{R}$ and where infinite time evolution preserves $p$, and (2) a space of bulk states $\mathcal{H}_{\rm bulk}$ with a basis $|\mathcal{P}\rangle$ with $\mathcal{P}>0$, where $\mathcal{P}$ is conserved. So the time evolution operator $\widehat{V}$ from the bulk to asymptotia is in fact a map from a smaller Hilbert space to a larger one, $\widehat{V}: \mathcal{H}_{\rm bulk} \to \mathcal{H}_{\rm asy}$. The natural (and correct) guess for $\widehat{V}$ is that it simply takes $\mathcal{P}$ to $p$. To show this consider the matrix element $\langle \Phi|\widehat{V}|\mathcal{P}\rangle$. This object is the de Sitter JT version~\cite{Maldacena:2019cbz, Cotler:2019nbi, Cotler:2019dcj} of the ``trumpet'' of AdS JT gravity~\cite{Saad:2019lba}, where the initial state fixes that the geometry is $ds^2 = -dt^2 + \mathcal{P}\cosh^2(t) dx^2$ with $t$ starting at some finite time. The JT path integral in this case reduces to a Schwarzian path integral at future infinity which depends on $\mathcal{P}$ and $\Phi$ with the result (in terms of normalized states $\langle \Phi|$ and $|\mathcal{P}\rangle$)
\beq
\label{E:Vbasischange1}
	\langle \Phi|\widehat{V}|\mathcal{P}\rangle \approx \frac{1}{\sqrt{2\pi}} \, e^{i \Phi \mathcal{P}}\,,
\eeq
which implies $\langle p|\widehat{V}|\mathcal{P}\rangle \approx \delta(p-\mathcal{P})$ as expected.

\begin{figure}[t]
\centering
\includegraphics[scale=.42]{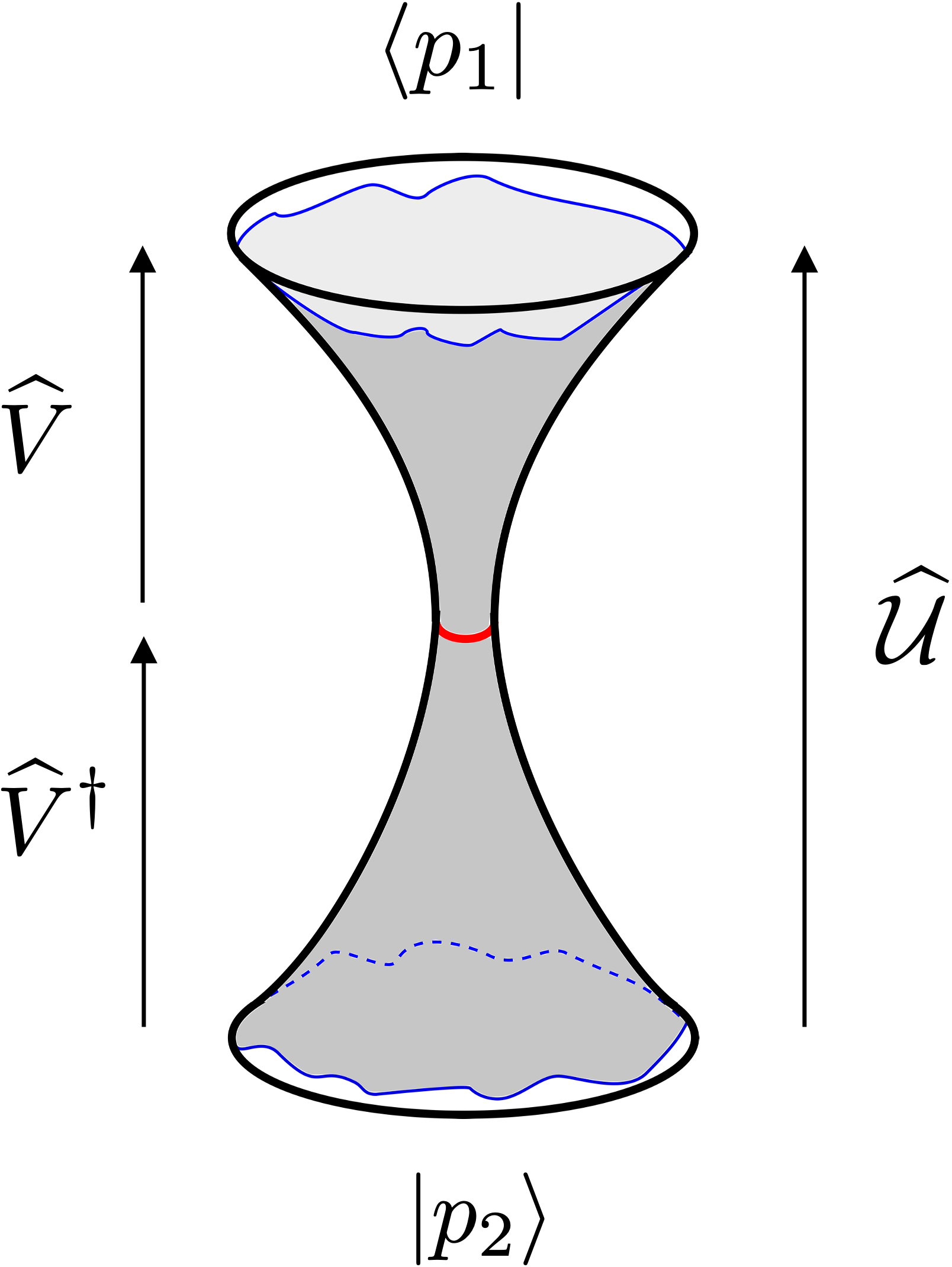}
\caption{The JT de Sitter S-matrix, starting in the state $|p_2\rangle$ and ending in the state $\langle p_1|$.  Time evolution from past infinity to the bottleneck is given by $\widehat{V}^\dagger$, and time evolution from the bottleneck to future infinity is given by $\widehat{V}$.  Since $\widehat{V}$ is an isometry, the total time evolution $\widehat{\mathcal{U}} = \widehat{V}\widehat{V}^\dagger$ is a projector. \label{Fig:Smatrix1}}
\end{figure}

The operator $\widehat{V}$ is therefore an isometry: the product $\widehat{V}^{\dagger}\widehat{V}$ acts as the identity on $\mathcal{H}_{\rm bulk}$ while $\widehat{V}\widehat{V}^{\dagger}=\widehat{\mathcal{U}}$ acts as a projector on $\mathcal{H}_{\rm asy}$. This is the main result of this paper.  A depiction of the de Sitter $S$-matrix can be seen in Figure~\ref{Fig:Smatrix1}.

Using the result~\eqref{E:Vbasischange1} we can reconstruct the bulk wavefunction of the no-boundary state.  In particular, since the Hartle-Hawking state is $|\text{HH}\rangle = \widehat{V} |\varnothing\rangle$, we have $\widehat{V}^\dagger |\text{HH}\rangle = |\varnothing\rangle$ and so
\beq
	\langle \mathcal{P}|\varnothing\rangle \approx  \delta'(\mathcal{P}-1)\,,
\eeq
which is supported on $\mathcal{P} = 1$ and is also clearly non-normalizable. This is consistent with the fact that the JT sphere partition function, expected on general grounds to be the norm of the no-boundary state, diverges~\cite{Mahajan:2021nsd}.


\section{Minisuperspace}

Now let us consider Einstein gravity with positive cosmological constant, where we put the metric in a temporal gauge so that $g_{\mu t} = \delta_t^\mu$.  This corresponds to the action
\beq
	S = \frac{1}{16\pi G} \int d^{d+1}x \sqrt{-g}\,(R-2\Lambda) + \int d^{d+1} x \,\sqrt{g_{\mathbb{S}^d}}\,\lambda^\mu(x) \left(g_{\mu t} - \delta_\mu^t\right) + (\text{bdy})\,,
\eeq
where we have included a Lagrange multiplier term to incorporate the gauge-fixing.  In that term, $g_{\mathbb{S}^d}$ is the round metric on $\mathbb{S}^d$.  We now consider a minisuperspace approximation (perhaps more appropriately called a `truncation') where the line element is
\beq
\label{E:miniSuperspace}
	ds^2 = - e^{2A(t)} dt^2 + e^{2B(t)}d\Omega_d^2
\eeq
and the Lagrange multiplier $\lambda^\mu$ is regarded as being solely a function of $t$.  As is well-known, the physics of the scale factor $B(t)$ is classically equivalent to a particle in a potential. Taking $\Lambda = \frac{d(d-1)}{2}$ so that the global de Sitter solution has unit de Sitter radius and defining $X(t) := e^{A(t)}$, $Y(t) := e^{\frac{d}{2}B(t)}$, $\lambda(t) := -\frac{d}{d-1}\,2\pi G\,\lambda^0(t)$, we have
\begin{align}
\begin{split}
	S &= - \frac{d-1}{d} \frac{\text{Vol}(\mathbb{S}^d)}{2\pi G} \int dt \left( \frac{1}{X}\,\frac{1}{2}\,\dot{Y}^2 - X\,V(Y) + \lambda(X-1)\right) + (\text{bdy})\,,
	\\
	V(Y) &= \frac{d^2}{8}\left( Y^{\frac{2(d-2)}{d}} - Y^{2}\right)\,.
\end{split}
\end{align}
The $\lambda$ equations of motion simply enforce $X = 1$, and the $Y$ equations of motion reduce to $\ddot{Y} = - V(Y)$ which describes a particle in a potential.  However, the $X$ equations of motion give $\lambda = \frac{1}{2}\,\dot{Y}^2 + V(Y)$ where we have set $X = 1$; but the right-hand side is simply the energy of $Y(t)$ and so $\lambda(t)$ must be constant.  Let us suggestively write $E = \lambda = \text{constant}$.  The standard solution comes from setting the Lagrange multiplier to zero, namely $E = 0$:
\begin{equation}
\label{E:Hconstraint1}
E = \frac{1}{2}\,\dot{Y}^2 + V(Y) = 0\,.
\end{equation}
This corresponds to imposing the Hamiltonian constraint (really the $g_{tt}$ equation of motion) which is that the total energy vanishes.  The above has the unique solution $Y^{\frac{4}{d}} = e^{2B} = \cosh^2(t)$, corresponding to global de Sitter space.

\begin{figure}[t]
\centering
\includegraphics[scale=.11]{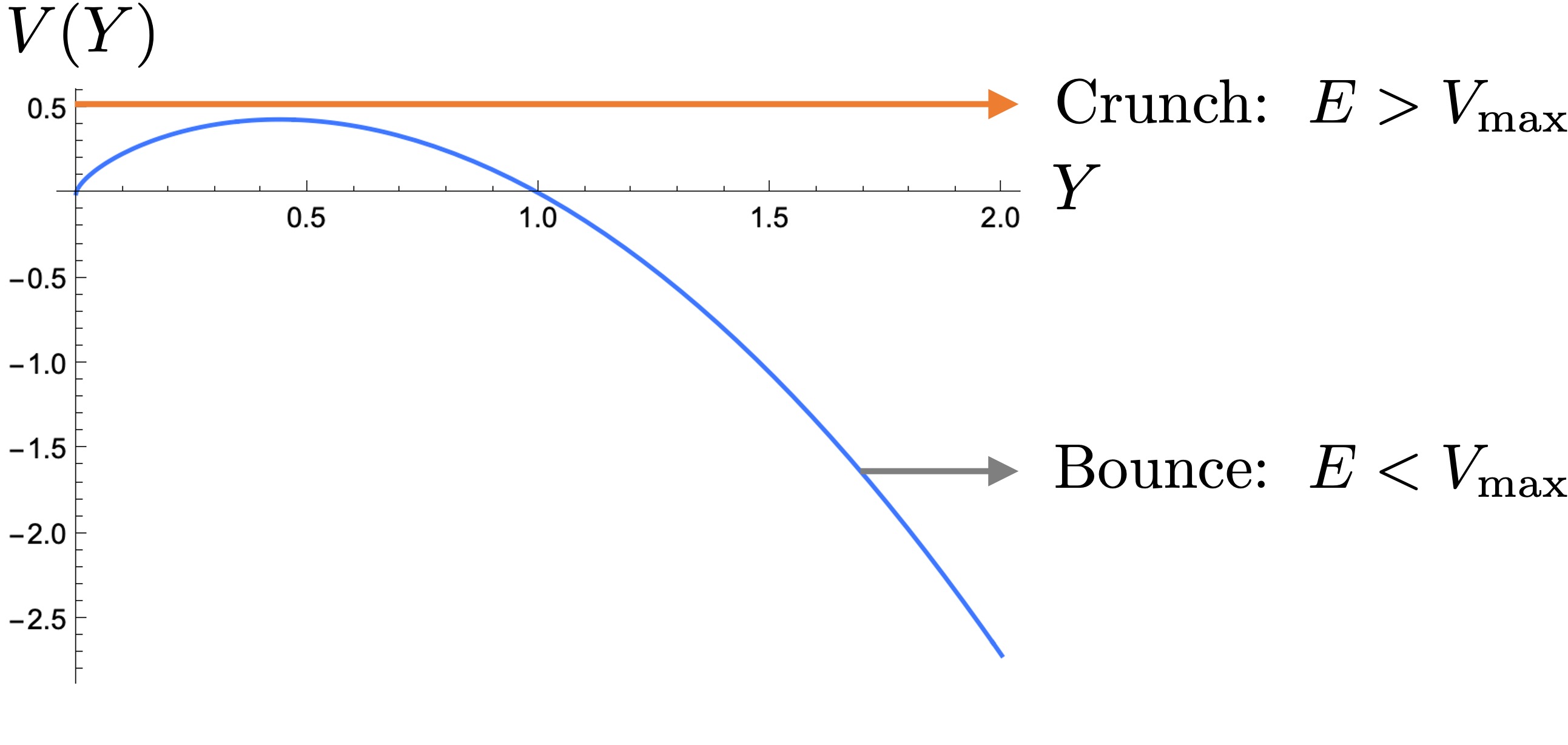}
\vspace{-.1cm}
\caption{Plot of potential for the minisuperspace equations.  For energies $E < V_{\max}$ the scale factor will reach a minimum size and bounce, whereas for $E > V_{\max}$ the warp factor will crunch to zero size.  \label{Fig:graph1}}
\end{figure}

But we need not set $E = 0$.  A non-zero $E$ corresponds to a non-zero Lagrange multiplier, which then contributes a constant forcing term to the $X$ the equations of motion.  That is, we now have
\begin{equation}
E = \frac{1}{2}\,\dot{Y}^2 + V(Y) \not = 0\,,
\end{equation}
corresponding to a $Y$ with energy $E$.  This is a minisuperspace analog of the gravitational constrained saddles in~\cite{Cotler:2020lxj} (see also the discussion of $\lambda$-solutions in~\cite{Cotler:2021cqa}), where the non-zero $E$ behaves as a forcing term.  A semiclassical quantization of our system then leads to states $|E\rangle$ of definite energy corresponding to geometries where $Y$ is fixed by $E$. We have plotted the potential $Y$ for $d=3$ (i.e.~in four spacetime dimensions) in Figure~\ref{Fig:graph1}.  For $E<V_{\rm max}$ there are ``scattering states'' corresponding to bounce cosmologies where the scale factor approaches some minimum size fixed by $E = V(Y)$, while for $E > V_{\max}$ there are states corresponding to crunch cosmologies where the trajectory ends with $e^{2B} \to 0$. A sum over smooth geometries will, semiclassically, keep only those states corresponding to bounce cosmologies with $E<V_{\rm max}$. This is reminiscent of our de Sitter JT analysis, where the bulk Hilbert space is only composed of states corresponding to non-singular geometries.

As an aside, we note that a similar restriction on the range of $E$ appears in the different but related context of two-sided Euclidean AdS wormholes~\cite{Cotler:2020ugk, Cotler:2020lxj, Cotler:2021cqa, Cotler:2022rud}.  In that context there is a lower bound on $E$ corresponding to the energy of the lightest black hole.  Some of the two-sided Euclidean AdS wormholes are related to our global de Sitter-type spacetimes by analytic continuation~\cite{Cotler:2020lxj}.

We have not undertaken a full analysis of whether the class of bouncing cosmologies with $E<V_{\rm max}$ are perturbatively stable against fluctuations in the full Einstein equations (i.e.~outside of the minisuperspace approximation). However, we have performed a partial check. Consider coupling in a massless scalar field $\zeta$ with action $-\frac{1}{2} \int d^{d+1} x \, \sqrt{-g}\,\partial_\mu \zeta \partial^\mu \zeta$. We have found that these backgrounds are stable against $s$-wave perturbations of the scale factor and of the scalar field.  To this end, let $\zeta = Z(t)$ so that our total action becomes
\begin{equation}
S_{\text{total}} = - \frac{d-1}{d} \frac{\text{Vol}(\mathbb{S}^d)}{2\pi G} \int dt \left( \frac{1}{X}\,\frac{1}{2}\,\dot{Y}^2 - X\,V(Y) + \lambda(X-1)\right) + \frac{1}{2}\,\text{Vol}(\mathbb{S}^d)\,\int dt \, \frac{Y^2}{X}\,\dot{Z}^2 + (\text{bdy})\,.
\end{equation}
As before the $\lambda$ equation of motion sets $X = 1$.  But now the $X$, $Y$, and $Z$ equations of motion, setting $X = 1$, are
\begin{align}
\label{E:EOMp1}
\lambda &= \frac{1}{2}\,\dot{Y}^2 + V(Y) - \frac{d}{d-1}\,\pi G\, Y^2\dot{Z}^2 \\
\label{E:EOMp2}
\ddot{Y} &= - V'(Y) + \frac{d}{d-1}\,\pi G \, Y \dot{Z}^2\\
\label{E:EOMp3}
\frac{d}{dt}(Y^2 \dot{Z}) &= 0\,.
\end{align}
The last equation shows us that $Q = Y^2 \dot{Z}$ is a conserved charge, and thus is constant.  The first two equations show us that $Y$ is a particle with energy $E = \lambda = \text{constant}$ in the potential
\begin{equation}
V_{\text{eff}}(Y, Q) := V(Y) - \frac{d}{d-1}\,\pi G\, \frac{Q^2}{Y^2}\,.
\end{equation}
This potential always has a global maximum $V_{\rm eff,max}$ at $Y_{\rm max}>0$. If $E<V_{\rm eff,max}$ the resulting spacetime is a smooth bounce. This holds for arbitrarily large $Q$, although for sufficiently large $Q$ the effective potential is always negative $V_{\rm eff,max}<0$. If we restrict ourselves to unforced geometries with $E=0$, then there is a finite range of $Q^2 \leq Q_c^2 = O(1/G)$ for which the spacetime is smooth. We can relate this incident scalar flux to a boundary stress tensor on the sphere in the far past. For $Q^2 = O(1/G)$ so that the scalar and cosmological contributions to the effective potential are comparable, that boundary stress tensor is $O(1/G)$. Holding the energy fixed we see that it takes a finite scalar perturbation in the infinite past to `over-close' the universe. This is a feature of global de Sitter-like spaces which differs from the physics of an inflating patch~\cite{Borde:2001nh}.


\section{Discussion}

Our work shows that the $S$-matrix need not be unitary in quantum gravity.  In de Sitter JT gravity this is a consequence of a mismatch between the bulk and asymptotic Hilbert spaces: initial conditions that correspond to crunching universes live in the Hilbert space of asymptotic states, but not in the ``code subspace'' of bulk states.  As a result, complete knowledge of bulk physics, even on arbitrarily large timescales, is not enough to deduce the de Sitter $S$-matrix.  We also find that evolution is trivial within the code subspace, with finite time evolution acting as the identity.  In our examples this breakdown of infinite-time unitary evolution and its replacement by a combination of projections and isometries is invisible in perturbation theory, but rather arises non-perturbatively.

Our JT example is particularly simple because it is model of pure gravity in low dimensions.
It would be natural to enrich our analysis by considering JT coupled to defects, worldlines, or conformal matter. In these settings we expect the code subspace of non-crunching geometries to be much richer, with an interplay between gravity and matter.

An important question is if some version of our results hold in more realistic models of quantum gravity.  The answer to this question is relevant for understanding the origins and ultimate fate of our universe, as we (presumably) live in a code subspace.  While we expect pure de Sitter quantum gravity in 2+1 dimensions, a model with no local degrees of freedom, to be rather similar to de Sitter JT gravity, the setting of Einstein gravity in 3+1 dimensions is less clear.  Our minisuperspace analysis offers a suggestion that the basic features of our work, a mismatch between bulk and asymptotic Hilbert spaces and isometric evolution, persists in more realistic settings.  One foreseeable question is whether the restriction to a sum over smooth geometries is realized in UV completions of de Sitter gravity, such as in string theory (if indeed a suitable stringy completion exists).  For instance, maybe certain singular metrics that are sensible in string theory ought to be included in the low-energy theory.

Perhaps a useful toy model to keep in mind is de Sitter JT gravity coupled to defects; here the universe can begin or end on a defect.  As such there is still a code subspace: it consists of all states which either evolve into a smooth geometry, or into a conical singularity that can be sourced by a defect.  However not all conical singularities are allowed since the set of defects is constrained.  This means that the code subspace encodes the `spectrum' of allowed singularities.

More broadly, we expect that the true Hilbert space of de Sitter quantum gravity is drastically smaller than the na\"{i}ve one indicated by semiclassical gravity.  In particular, holographic arguments suggest that the actual dimension is non-perturbatively finite.  Our findings are a first step in this direction, where we can already see the pruning of the bulk Hilbert space in the low-energy effective description.

\subsection*{Acknowledgements}

We would like to thank V.~Hubeny, L.~Iliesiu, and A.~Strominger for enlightening discussions. JC is supported by a Junior Fellowship from the Harvard Society of Fellows, as well as in part by the Department of Energy under grant DE-SC0007870.  KJ is supported in part by an NSERC Discovery Grant.

\appendix
\section{Refined JT/matrix model dictionary}

We begin by reviewing the path integral for the Schwarzian theory with a varying dilaton. We require
\begin{align}
\begin{split}
	Z_0[\varphi] &= \int \frac{[df]}{PSL(2;\mathbb{R})} \exp\left( \frac{1}{\pi}\int_0^{2\pi}d\tau \, e^{\varphi(\tau)} \left( \{f(\tau),\tau\} + \frac{1}{2}f'(\tau)^2\right)\right) \,,
	\\
	 Z_{b^2}[\varphi] & = \int \frac{[df]}{U(1)} \exp\left( \frac{1}{\pi}\int_0^{2\pi} d\tau\,\underbrace{e^{\varphi(\tau)}\left(\{f(\tau),\tau\} - \frac{b^2}{2}f'(\tau)^2\right)}_{\equiv \mathcal{T}(\tau)}\right)\,,
\end{split}
\end{align}
where $e^{\varphi(\tau)}$ is the renormalized dilaton profile. The first integral was evaluated in Appendix C of~\cite{Stanford:2017thb} when the dilaton profile takes the form $e^{-\varphi(\tau)} = g^2 h'(\tau)$ for $g^2>0$ and is $h(\tau)$ a reparameterization of the circle. The second integral can be similarly evaluated.

Here we expand slightly on that result. It turns out that each of these integrals has a saddle point approximation; the classical equation of motion reads\footnote{We are describing the equations of motion for both Schwarzian models simultaneously, with the understanding that we substitute $b^2=-1$ to study the exceptional model where $f(\tau)\in \faktor{\text{Diff}(\mathbb{S}^1)}{PSL(2;\mathbb{R})}$.}
\beq
\label{E:schwarzianEOM}
	\frac{d\mathcal{T}(\tau)}{d\tau}  + \varphi'(\tau) \mathcal{T}(\tau)+  \frac{d^3e^{\varphi(\tau)}}{d\tau^3}=0\,,
\eeq
which has a solution $f(\tau) = \sigma(\tau)$ obeying
\beq
	\sigma'(\tau) = \frac{e^{-\varphi(\tau)}}{\frac{1}{2\pi}\int_0^{2\pi}d\tau\,e^{-\varphi(\tau')}}\,.
\eeq
This is a genuine reparameterization with $\sigma(\tau+2\pi)=\sigma(\tau)+2\pi$ and $\sigma'\geq 0$ provided that $e^{-\varphi}\geq 0$. Note that $e^{-\varphi(\tau)} = \Phi^{-1} \sigma'(\tau)$ with $\Phi^{-1} = \frac{1}{2\pi}\int_0^{2\pi} d\tau \,e^{-\varphi(\tau)}$, so that we can regard $\sigma(\tau)$ as $h(\tau)$ and $g^2$ as $\Phi^{-1}$. 

Because $\sigma(\tau)$ is a reparameterization of the circle, we can invert the relation between $\sigma$ and $\tau$ to obtain $\tau(\sigma)$. Changing variables from $f(\tau)$ to $f(\sigma)$ and using the chain rule for the Schwarzian derivative
\beq
	\{f(g(\tau)),\tau\} = \{g(\tau),\tau\} + \{f(g),g\} g'(\tau)^2\,,
\eeq
we have
\begin{align}
\label{E:changeOfTime}
	\frac{1}{\pi}\int_0^{2\pi} d\tau\,e^{\varphi} \left( \{f(\sigma(\tau)),\tau\} - \frac{b^2}{2}f'(\tau)^2\right) & =\frac{\Phi}{\pi} \int_0^{2\pi} d\sigma\left( \{f(\sigma),\sigma\} - \frac{b^2}{2}f'(\sigma)^2\right) + \frac{1}{2\pi} \int_0^{2\pi} d\tau\,e^{\varphi(\tau)}\varphi'(\tau)^2\,.
\end{align}
So by this change of coordinates we can write the Schwarzian action with a varying dilaton as a Schwarzian action with a constant dilaton $\varphi = \ln \Phi$ plus a term that only depends on the background fields and which therefore factors out of the path integral. 

The spaces $\faktor{\text{Diff}(\mathbb{S}^1)}{PSL(2;\mathbb{R})}$ and $\faktor{\text{Diff}(\mathbb{S}^1)}{U(1)}$ are symplectic with symplectic form
\beq
	\omega = \int_0^{2\pi} d\tau\,\left( \frac{df'(\tau) \wedge df''(\tau)}{f'(\tau)^2} + b^2 df(\tau) \wedge df'(\tau)\right)\,.
\eeq
(Again we simply substitute $b^2=-1$ when considering $\faktor{\text{Diff}(\mathbb{S}^1)}{PSL(2;\mathbb{R})}$.) The measure over $f(\tau)$ is the symplectic measure associated with $\omega$. This symplectic form is invariant under the change of coordinates from $\tau$ to $\sigma$, and so the measure is too. We then have
\beq
	Z_{b^2}[\varphi] = e^{\frac{1}{2\pi}\int_0^{2\pi} d\tau\,e^{\varphi}\varphi'^2} \int \frac{[df]}{U(1)} e^{\frac{\Phi}{\pi} \int_0^{2\pi} d\sigma\left( \{f(\sigma),\sigma\} - \frac{b^2}{2} \left( \frac{df}{d\sigma}\right)^2\right)}=e^{\frac{1}{2\pi}\int_0^{2\pi} d\tau\,e^{\varphi}\varphi'^2}Z_{b^2}[\ln \Phi]\,,
\eeq
and similarly for $Z_0$.

The Schwarzian path integrals with constant dilaton are known by localization~\cite{Stanford:2017thb}. Using those we then have
\begin{align}
\begin{split}
\label{E:generalSchwarzianZ}
	Z_0[\varphi] & = \mathcal{N}_0 \exp\left( \frac{1}{2}\int_0^{2\pi} d\tau\,e^{\varphi}\varphi'^2 +  \Phi+ \frac{3}{2}\ln \Phi\right)\,,
	\\
	Z_{b^2}[\varphi]& = \mathcal{N} \exp\left( \frac{1}{2}\int_0^{2\pi} d\tau\,e^{\varphi}\varphi'^2 -  b^2\Phi + \frac{1}{2}\ln \Phi\right)\,,
\end{split}
\end{align}
where $\mathcal{N}_0$ and $\mathcal{N}$ are normalization constants that depend on the choice of regularization. One such choice is $\mathcal{N}_0= \mathcal{N}=\frac{1}{\sqrt{2\pi}}$.

Another way to arrive at these results is through Ward identities. In the quantum version of the Schwarzian theory the equation of motion~\eqref{E:schwarzianEOM} is promoted to an operator identity, with
\beq
\label{E:schwarzianWard}
	\frac{d\langle \mathcal{T}(\tau)\rangle}{d\tau} + \varphi'(\tau)\langle \mathcal{T}(\tau)\rangle +\frac{d^3e^{\varphi(\tau)}}{d\tau^3} = 0\,, \qquad \langle \mathcal{T}(\tau)\rangle =\pi \frac{\delta \ln Z}{\delta \varphi(\tau)}\,.
\eeq
This Ward identity is in fact also the diffeomorphism Ward identity associated with the Schwarzian theory. Coupling the Schwarzian model to an external metric $g_{\tau\tau}(\tau)$, the stress tensor $\langle T^{\tau\tau}\rangle = -\frac{2}{\sqrt{g}}\frac{\delta \ln Z}{\delta g_{\tau\tau}}$ obeys the Ward identity $D_{\nu}\langle T^{\mu\nu}\rangle = -D^{\mu}\varphi \langle \mathcal{T}\rangle/\pi$ with $D_{\mu}$ the covariant derivative. A simple computation reveals that the stress tensor with $g_{\tau\tau}=1$ is $\pi T^{\tau\tau} = \mathcal{T}- \frac{d^2e^{\varphi}}{d\tau^2}$, so that the diffeomorphsim Ward identity coincides with~\eqref{E:schwarzianWard}. In any case the Ward identity~\eqref{E:schwarzianWard}, viewed as a differential equation for $\langle \mathcal{T}\rangle$, has a simple solution
\beq
	\langle \mathcal{T}(\tau)\rangle = -\frac{e^{\varphi(\tau)}}{2}(\varphi'(\tau)^2+2\varphi''(\tau)) - \frac{\mathcal{C}}{2}\,e^{-\varphi(\tau)}\,,
\eeq
for an integration constant $\mathcal{C}$. We can integrate this functional variation to obtain 
\beq
	\ln Z[\varphi] = \frac{1}{2\pi}\int_0^{2\pi} d\tau\,e^{\varphi}\varphi'^2 + F(\Phi^{-1})\,, \qquad \mathcal{C} = \frac{\partial F}{\partial \Phi^{-1}}\,,
\eeq
where we have used that $\delta\Phi^{-1} = -\frac{1}{2\pi} \int_0^{2\pi}d\tau\, e^{-\varphi(\tau)}\delta\varphi(\tau)$. Setting the dilaton to be constant so that $\varphi = \ln \Phi$ and matching to the partition function with constant dilaton we have $F(\Phi) = \ln Z[\ln \Phi]$ and so
\beq
	\ln Z[\varphi] = \frac{1}{2\pi}\int_0^{2\pi} d\tau\,e^{\varphi}\varphi'^2 + \ln Z[\ln \Phi]\,,
\eeq
coinciding with~\eqref{E:generalSchwarzianZ}.

With the path integrals~\eqref{E:generalSchwarzianZ} in hand we can evaluate the most general amplitude $Z_{g,n}$ for JT gravity on Euclidean $R=-2$ surfaces of genus $g$ with $n$ asymptotic boundaries. We parameterize each boundary circle with $\tau \sim \tau+2\pi$ and on the $i$th boundary have a renormalized dilaton $e^{\varphi_i}$. Then by the same logic as~\cite{Stanford:2017thb} the JT amplitudes with a varying dilaton are simply related to those with a constant dilaton by
\beq
	Z_{g,n}[\{\varphi_i\}] = \exp\left( \frac{1}{2\pi} \sum_{i=1}^n \int_0^{2\pi}  d\tau\,e^{\varphi_i(\tau)}\varphi_i'(\tau)^2\right) Z_{g,n} [\{\ln \Phi_i\}]\,,
\eeq
where the amplitude $Z_{g,n}[\{\ln \Phi_i\}]$ is given in~\cite{Saad:2019lba}. So the effect of a dilaton varying on the $i$th boundary is to simply multiply the standard JT amplitude by a local prefactor depending on that boundary.

The constant dilaton amplitudes $Z_{g,n}[\{\ln \Phi_i\}]$ are known to be equal to quantities computed from a particular double scaled matrix model~\cite{Saad:2019lba}. The matrix model involves a single Hermitian matrix $H$ with averages
\beq
	\langle f(H)\rangle_{\rm MM} = \int dH \,e^{-\text{tr}(V(H))} f(H)\,,
\eeq
for a suitable potential $V(H)$. The matrix model averages at hand have a genus expansion. The dictionary reads, with some convention for $H$
\beq
	Z_{g,n}[\{\ln \Phi_i\}] = \left\langle \text{tr}\!\left( e^{-\Phi_1^{-1} H}\right) \cdots \text{tr}\!\left( e^{-\Phi_n^{-1} H}\right)\right\rangle_{\text{ MM,\,conn},\,g}\,,
\eeq
where the right-hand side refers to the genus-$g$ term in the connected part of the matrix model average. That is, with constant dilaton, an asymptotic boundary with constant dilaton corresponds to an insertion of $\text{tr}\!\left( e^{-\Phi^{-1} H}\right)$ into the matrix model average. It then follows that we continue to have a duality between JT amplitudes and matrix model averages when in JT we have a varying dilaton. An asymptotic boundary with varying dilaton corresponds to an insertion of
\beq
\includegraphics[scale=.32, valign = c]{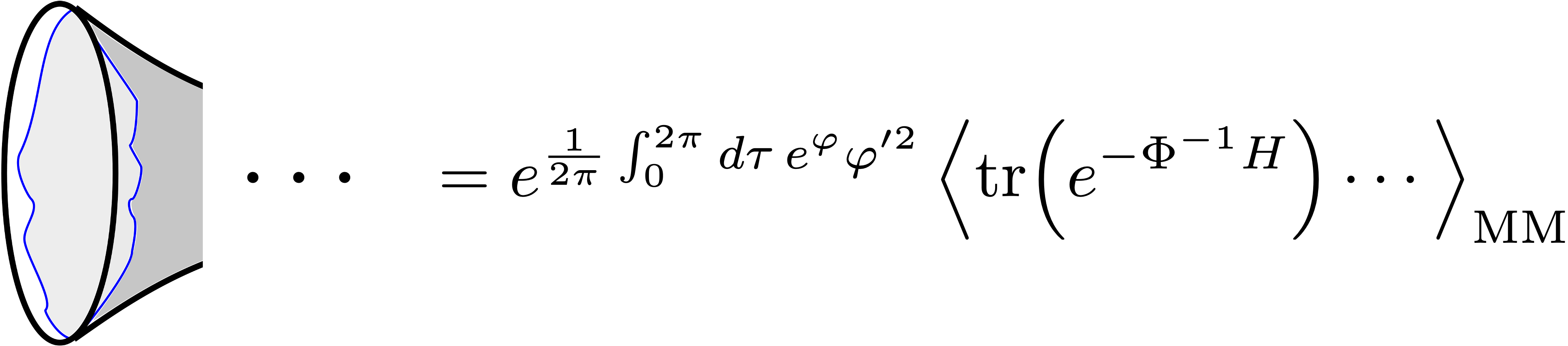}
\eeq
into the matrix model average.

Given the dictionary with a constant dilaton, we can understand the dictionary with a varying dilaton intuitively as follows. As we mentioned above, the stress tensor with $g_{\tau\tau}=1$ is given by $\pi T^{\tau\tau} = \mathcal{T} - \frac{d^2e^{\varphi}}{d\tau^2}$. In the Schwarzian theory this corresponds to a time-dependent Hamiltonian
\beq
	\pi \mathcal{H}(\tau) = -\mathcal{T}(\tau) +\frac{d^2e^{\varphi}}{d\tau^2} = - e^{\varphi(\tau)}\left( \{f(\tau),\tau\} - \frac{b^2}{2}f'(\tau)^2\right) + \frac{d^2e^{\varphi}}{d\tau^2}\,.
\eeq
Note that the time-dependence appears in two ways: through an additive contribution $\frac{d^2e^{\varphi}}{d\tau^2}$ independent of the quantum fields, and through an overall multiplicative factor $e^{\varphi(\tau)}$. Thus the Euclidean evolution operator requires no time-ordering and the Schwarzian path integral, if there was a Hilbert space interpretation, would be understood as
\beq
	Z[\varphi] = \text{tr}\!\left( e^{-\int_0^{2\pi} d\tau\, \mathcal{H}(\tau)}\right)=\text{tr}\!\left( e^{-\frac{1}{\pi}\int_0^{2\pi} d\tau\,e^{\varphi}\left( \{f(\tau),\tau\} - \frac{b^2}{2}f'(\tau)^2\right)}\right)\,.
\eeq
But, by the change of coordinate from $\tau$ to $\sigma$ and~\eqref{E:changeOfTime}, we then have
\begin{align}
\begin{split}
	Z[\varphi] = \text{tr}\!\left( e^{-\int_0^{2\pi}d\tau\,\mathcal{H}(\tau)}\right) &= e^{\frac{1}{2\pi}\int_0^{2\pi} d\tau\,e^{\varphi}\varphi'^2} \text{tr}\!\left( e^{\frac{\Phi}{\pi}\int_0^{2\pi}d\sigma \left( \{f(\sigma),\sigma\} - \frac{b^2}{2}f'(\sigma)^2\right)}\right)
	\\
	& =  e^{\frac{1}{2\pi}\int_0^{2\pi} d\tau\,e^{\varphi}\varphi'^2}\text{tr}\!\left( e^{-\Phi^{-1} H}\right)\,.
\end{split}
\end{align}

\section{Some details about de Sitter JT gravity}
\label{App:details2}

The goal of this appendix is to summarize the key facts about de Sitter JT gravity we use in the main text. Many of these facts are already contained in~\cite{Cotler:2019dcj}.

The action of de Sitter JT gravity is
\beq
	S = \frac{S_0}{4\pi}\left( \int d^2x \sqrt{-g}\,R - 2 \int dx \sqrt{\gamma}\,K\right) + \int d^2x \sqrt{-g}\,\phi(R-2) -2 \int dx\sqrt{\gamma} \,\phi(K-1)\,,
\eeq
Integrating out the dilaton enforces that the metric has constant curvature. We focus on the case where spacetime fills in a circle at future infinity, and when spacetime connects a circle in the far past to a circle in the far future. The first case is related to the JT version of the no-boundary state, and the second to infinite-time transition amplitudes. In either case the constant curvature condition allows us to completely fix the metric.

When there is a single circle in the far future, the spacetime is
\beq
\label{E:metricHH}
	ds^2 = -dt^2 + \cosh^2(t) dx^2\,,
\eeq
where $t$ is taken to be a complex time contour with two segments: the first is $t>0$, connecting a bottleneck at $t=0$ with an asymptotically large circle in the future; the second is $t=i\tau$ with $\tau \in [0,\pi/2)$, on which the line element reads (recall that $x\sim x+2\pi$)
\beq
	ds^2 = d\tau^2 + \cos^2(\tau)dx^2\,.
\eeq
The real time segment is half of global dS$_2$, while the imaginary time segment is a Euclidean hemisphere, which we interpret as preparing the no-boundary state $|\varnothing\rangle$. See Figure~\ref{Fig:HH1} in the main text.  The total spacetime has the topology of a disk. Having completely fixed the spacetime metric one might think that there is nothing left; this is not the case, and there is a single boundary degree of freedom remaining. One can think of this degree of freedom as corresponding to a freedom in picking the future boundary in such a way as to obey the future dS$_2$ boundary conditions
\begin{align}
\begin{split}
	ds^2 & = -dt^2 + (e^{2t} + O(1))dx^2\,,
	\\
	\phi & = \frac{1}{2\pi}\,e^{t + \varphi(x)} + O(1)\,,
\end{split}
\end{align}
or in terms of large diffeomorphisms acting on the metric~\eqref{E:metricHH}. Either way, one finds a boundary degree of freedom with a Schwarzian effective action. See~\cite{Maldacena:2019cbz, Cotler:2019nbi} for the details. The final result is that the boundary degree of freedom can be taken to be a reparameterization $f(x)$ of the circle at infinity with an action
\beq
	S =-i S_0 +  \frac{1}{\pi}\int_0^{2\pi} dx\,e^{\varphi(x)}\left( \{f(x),x\} + \frac{1}{2}f'(x)^2\right)\,.
\eeq
The topological term proportional to $S_0$ arises from the Euclidean hemisphere. Strictly speaking $f(x)$ is an element of the quotient space $\faktor{\text{Diff}(\mathbb{S}^1)}{PSL(2;\mathbb{R})}$. Because it is a reparameterization it obeys
\beq
	f(x+2\pi) = f(x)+2\pi\,, \qquad f'(x)>0\,,
\eeq
while the quotient means that we identify
\beq
	\tan\left( \frac{f(x)}{2}\right) \sim \frac{a \tan\left( \frac{f(x)}{2}\right) + b}{c\tan\left(\frac{f(x)}{2}\right)+d}\,, \qquad ad-bc=1\,,
\eeq
arising from the $PSL(2;\mathbb{R})$ isometry of the spacetime. The JT path integral reduces to one over this Schwarzian degree of freedom,
\beq
	Z_{\rm disk} = e^{S_0} \int \frac{[df]}{PSL(2;\mathbb{R})} \exp\left( \frac{i}{\pi}\int_0^{2\pi} dx\,e^{\varphi(x)}\left( \{f(x),x\} + \frac{1}{2}f'(x)^2\right)\right)\,.
\eeq
As long as $e^{\varphi}$ is everywhere positive, or everywhere negative, this integral can be evaluated as an analytic continuation of our results of the last Subsection with 
\beq
	Z_{\rm disk} =\frac{(i\Phi)^{\frac{3}{2}}}{\sqrt{2\pi}}\,e^{i\Phi + i \mathcal{S}[\varphi]}\,, \qquad \mathcal{S}[\varphi] = \frac{1}{2\pi}\int_0^{2\pi} dx\,e^{\varphi(x)}\varphi'(x)^2\,.
\eeq

Now consider the problem where spacetime smoothly connects an asymptotic circle in the far past to one in the far future. The spacetime topology is a cylinder; unlike for the disk, there is now a moduli space of constant curvature metrics to integrate over. The line element can be fixed as
\beq
	ds^2 = -dt^2 + \alpha^2 \cosh^2(t)(dx+f'(t)dt)^2\,,
\eeq
where $\alpha>0$ is one modulus and $f(t)$ encodes the other. By $f(t)$ we mean a large diffeomorphism $x\to x+f(t)$ where $f$ approaches constant values in the far past and future. We can take $f(t) = \gamma \Theta(t)$ where $\gamma$ is the other modulus with $\gamma \sim \gamma+2\pi$. In addition to the moduli there are boundary degrees of freedom weighted by a Schwarzian effective action. The total amplitude reads
\begin{align}
\begin{split}
\label{E:Zcylinder}
	Z_{\rm cylinder} &= \frac{1}{2\pi} \int_0^{\infty} d\alpha^2 \,\int_0^{2\pi} d\gamma \,Z_T[\varphi_1] Z_T[\varphi_2]\,,
	\\
	Z_T[\varphi]& = \int \frac{[df]}{U(1)} \exp\left( \frac{i}{\pi}\int_0^{2\pi}dx\,e^{\varphi(x)}\left( \{f(x),x\} + \frac{\alpha^2}{2}f'(x)^2\right)\right)\,,
\end{split}
\end{align}
where $\varphi_1$ is the dilaton profile on the future circle and $\varphi_2$ the profile on the past circle. The moduli space measure is the Weil-Petersson measure on the space of constant curvature metrics and $f(x)$ is an element of $\faktor{\text{Diff}(\mathbb{S}^1)}{U(1)}$ with $f(x)\sim f(x) + a$ with $a\sim a+2\pi$. The Schwarzian path integral exists for $e^{\varphi}$ always positive or negative, and is given by the continuation of our results in the previous Appendix as
\beq
	Z_T = \sqrt{\frac{i\Phi}{2\pi}}\, e^{i \alpha^2 \Phi + i \mathcal{S}[\varphi]}\,.
\eeq
To render the integral over $\alpha^2$ convergent we introduce an $i\epsilon$ prescription, sending $\Phi \to \Phi + i \epsilon$, so that the cylinder amplitude reads
\beq
	Z_{\rm cylinder} =\frac{1}{2\pi} \frac{\sqrt{\Phi_1 \Phi_2}}{\Phi_1+\Phi_2 + i \epsilon} \,e^{i\mathcal{S}[\varphi_1]+i \mathcal{S}[\varphi_2]}\,.
\eeq

The disk and cylinder partition functions are unnormalized transition amplitudes with\footnote{Here and later the $\approx$ refers to the fact that we are not including genus corrections.}
\beq
	Z_{\rm disk} \approx \langle e^{\varphi}|\text{HH}\rangle = \langle e^{\varphi}|\widehat{V}|\varnothing\rangle \,,
\eeq
where the out state $\langle e^{\varphi}|$ has not yet been normalized. To extract normalized amplitudes we must first find the inner product of asymptotic states. The inner product of future asymptotic states proceeds as explained in~\cite{Cotler:2019dcj} and as described in the main text, and closely resembles the cylinder partition function. There are two minor differences. First, the modulus $\alpha^2$ is now valued on the real line, since in the asymptotic region the line element becomes
\beq
	ds^2 = -dt^2 + \left( e^{2t} + \frac{\alpha^2}{2}  + O(e^{-2t})\right)dx^2\,,
\eeq
which is non-singular for any value of $\alpha^2$. Second, we must account for the fact that the extrinsic curvature of the circle that prepares the ket is negative, so that the action of the Schwarzian mode on it is minus that in~\eqref{E:Zcylinder}. The net result is that the inner product is
\beq
	\langle e^{\varphi_1}|e^{\varphi_2}\rangle \approx \frac{1}{2\pi}\int_{-\infty}^{\infty}d\alpha^2 \int_0^{2\pi} d\gamma\,Z_T[\varphi_1]Z_T^*[\varphi_2]  = \sqrt{\Phi_1\Phi_2}\,\delta(\Phi_1-\Phi_2)\,e^{i\mathcal{S}[\varphi_1]-i \mathcal{S}[\varphi_2]}\,,
\eeq
which we advertised in the main text.

In the far past we perform a similar computation and find that that the inner product is the complex conjugate of that in the future, essentially because now the extrinsic curvatures of the circles preparing the ket and bra are flipped; the circle preparing the ket is further in the past and has positive extrinsic curvature, while the one preparing the bra has negative extrinsic curvature. 

To simplify matters we then define the past asymptotic state $|e^{\varphi}\rangle$ to be the state prepared by a constant renormalized dilaton $-e^{\varphi(x)}$, so that the inner product of past asymptotic states coincides with that of future asymptotic states. The unnormalized cylinder amplitude is then
\beq
	Z_{\rm cylinder} = \frac{i}{2\pi} \frac{\sqrt{\Phi_1\Phi_2}}{\Phi_1-\Phi_2+i\epsilon}\,e^{i\mathcal{S}[\varphi_1]-i\mathcal{S}[\varphi_2]}\approx \langle e^{\varphi_1}|\,\widehat{\mathcal{U}}\,|e^{\varphi_2}\rangle \,.
\eeq
As explained in the main text we can ``straighten out'' the dilaton, exchanging the states $|e^{\varphi}\rangle$ for those $|\Phi\rangle$ with constant dilaton (and same zero mode of $e^{-\varphi}$), so that the inner product is simply
\beq
	\langle \Phi_1|\Phi_2\rangle \approx \sqrt{\Phi_1\Phi_2}\,\delta(\Phi_1-\Phi_2)\,,
\eeq
and then rescale the asymptotic states $|\Phi\rangle \to \frac{|\Phi\rangle}{\sqrt{\Phi}}$ so as to be canonically normalized. The normalized infinite time transition amplitude is then
\beq
	\langle \Phi_1|\,\widehat{\mathcal{U}}\,|\Phi_2\rangle \approx \frac{i}{2\pi}\frac{1}{\Phi_1-\Phi_2+i\epsilon}\,,
\eeq
as claimed in the main text.  Instead working in a canonically conjugate momentum basis, we find
\begin{equation}
\label{E:pSmatrix1}
\langle p_1 | \, \widehat{\mathcal{U}} \, |p_2\rangle \approx \Theta(p_1) \delta(p_1 - p_2)\,.
\end{equation}


\bibliography{refs}
\bibliographystyle{JHEP}

\end{document}